\documentclass[preprint2]{aastex}
\shorttitle{LS~I~+61~303} 
\shortauthors{Grundstrom et al.} 
\begin{document} 
 
\accepted{19 October, 2006}
 
\title{Joint H$\alpha$ and X-Ray Observations of Massive X-Ray Binaries. II. 
The Be X-ray Binary and Microquasar LS~I~+61~303}  
 
\author{E. D. Grundstrom, S. M. Caballero-Nieves, D. R. Gies\altaffilmark{1}, \\ 
 W. Huang\altaffilmark{1,2}, M. V. McSwain\altaffilmark{1,3,4}, S. E. Rafter, 
 R. L. Riddle\altaffilmark{1,5}, \\ S. J. Williams, and 
 D. W. Wingert\altaffilmark{1}} 
\affil{Center for High Angular Resolution Astronomy and \\ 
 Department of Physics and Astronomy,\\ 
 Georgia State University, P. O. Box 4106, Atlanta, GA 30302-4106; \\ 
 erika@chara.gsu.edu, scaballero@chara.gsu.edu, gies@chara.gsu.edu,   
 wenjin@astro.caltech.edu, mcswain@astro.yale.edu, rafter@chara.gsu.edu,  
 riddle@tmt.org, swilliams@chara.gsu.edu, wingert@chara.gsu.edu} 
 
\altaffiltext{1}{Visiting Astronomer, Kitt Peak National Observatory, 
National Optical Astronomy Observatory, operated by the Association 
of Universities for Research in Astronomy, Inc., under contract with 
the National Science Foundation.} 
\altaffiltext{2}{Current address: Department of Astronomy, 
 California Institute of Technology, MS 105-24, 
 Pasadena, CA 91125} 
\altaffiltext{3}{Current Address: Astronomy Department, 
Yale University, New Haven, CT 06520-8101} 
\altaffiltext{4}{NSF Astronomy and Astrophysics Postdoctoral Fellow} 
\altaffiltext{5}{Current Address: Thirty Meter Telescope, 
 2632 E. Washington Blvd., Pasadena, CA 91107} 
 
\slugcomment{Accepted to ApJ} 
\paperid{65785} 

 
\begin{abstract} 
 
We present the results of an H$\alpha$ monitoring campaign  
on the BeXRB and microquasar system LS~I~+61~303.    
We use radial velocity measurements of \ion{He}{1} lines  
in our spectra to re-evaluate the orbital elements  
and to better establish the time of periastron.   
We list equivalent widths and other parameters for the  
H$\alpha$ emission line and discuss the orbital phase  
related variations observed.  We call attention to a  
dramatic episode of emission weakening that occurred in  
less than a day that probably resulted from exposure to  
a transient source of ionizing radiation.  We argue that  
the increase in H$\alpha$ and X-ray emission following  
periastron probably results from the creation of an  
extended density wave in the disk created by tidal forces.  
We also discuss estimates of the size of the disk from 
the H$\alpha$ equivalent width measurements, and  
we suggest that the disk radius from the average  
equivalent width corresponds to a resonant truncation  
radius of the disk while the maximum equivalent width 
corresponds to a radius limited by the separation 
of the stars at periastron.   
We note that a nearby faint companion is probably an  
unrelated foreground object. 
 
\end{abstract} 
 
 
 
\setcounter{footnote}{5} 
 
\section{Introduction}                              
 
The remarkable binary system LS~I~+61~303 is one of the  
best studied microquasars in the sky \citep{mas04}.   It consists of a  
rapidly rotating Be star surrounded by a dense circumstellar  
disk and a compact companion that is a source of both  
orbital phase modulated X-ray \citep{par97,lea01,wen06} and  
radio flux \citep{tay92,gre02}.   High angular resolution 
observations suggest that the system has precessing,  
relativistic radio jets \citep{mea04}, and these jets may 
be the source of variable and very high energy, 
$\gamma$-ray emission \citep{alb06}.  The companion  
is generally considered to be a magnetic neutron star,  
but it may be a black hole \citep{pun99}.    The orbital  
parameters were recently reanalyzed by \citet{cas05}  
who confirmed the high eccentricity of the system and  
discussed the possible mass range.  Their work shows  
that the orbital phases of the X-ray and radio maxima  
do not occur at periastron but are curiously delayed  
by a significant fraction of the orbital period.  
The disk that feeds the companion is a strong source of  
H$\alpha$ emission, and there is a well established  
record of the emission line variations over the past decade 
\citep{par94,zam99,liu00,zam01,liu05}.  \citet{zam99}  
demonstrated that some of the emission line properties  
vary with the orbital period, presumably as the result  
of the companion's varying gravitational pull.  
 
Here we present the results of a new observational effort  
on the H$\alpha$ variations in LS~I~+61~303 that we  
launched to investigate the co-variations with the  
X-ray flux as recorded by the All-Sky Monitor (ASM) instrument  
aboard the {\it Rossi X-ray Timing Explorer (RXTE)} satellite.  
It follows similar investigations of Cyg~X-1 \citep{gie03},   
LS~I~+65~010 \citep{gr06a}, and HDE~245770 = A~0535+26 
and X~Per \citep{gr06b}.  We first describe  
radial velocity measurements and a re-evaluation of the orbital  
elements in \S2.  Then we turn in \S3 to the temporal  
variations observed in the H$\alpha$ profile on rapid,  
orbital, and long term timescales.  We discuss the results  
in the context of the disk truncation model of \citet{oka02}  
in \S4.   
 
 
\section{Radial Velocities and Orbital Elements}    
 
We obtained a set of 100 spectra of LS~I~+61~303 between  
1998 August and 2000 December.  The observations were made with  
the Kitt Peak National Observatory 0.9~m coud\'{e} feed  
telescope during six runs.  We used two spectrograph  
arrangements to record the red spectral region around H$\alpha$ 
with resolving powers of $R=\lambda / \triangle \lambda = 4100$  
and 9500.  The details about the spectra and their reduction   
are given in a companion paper on LS~I~+65~010 \citep{gr06a}.  
We also recorded the spectrum of a nearby visual companion, 
and we discuss its properties in the Appendix. 
 
All the spectra record the H$\alpha$ emission line and the 
\ion{He}{1} $\lambda 6678$ feature.  The \ion{He}{1} $\lambda 6678$ 
line appears in most spectra as a weak and narrow absorption 
core flanked by emission wings (see Fig.~2 below).  We suspect  
that this line is a composite of a partial photospheric absorption 
line with wing emission and central absorption that originates 
in the inner disk.  We assume that the radial velocity variations 
of this composite profile represent the motion of the Be star  
since the line formation probably occurs very close to the  
photosphere of the Be star.  We note in passing that the  
half-width at half-minimum in the absorption core of  
\ion{He}{1} $\lambda 6678$ is 101 km~s$^{-1}$ in most of  
our spectra.  This is similar to the projected rotational  
velocity $V\sin i = 113$ km~s$^{-1}$ reported by \citet{cas05},  
but much smaller than that found earlier by \citet{hut81}  
of $V\sin i = 360 \pm 25$ km~s$^{-1}$.  This suggests that  
some of the blue spectral lines measured by \citet{cas05}  
may also have been affected by weak emission in the line  
wings that made the absorption cores appear more narrow.  
 
We measured radial velocities by cross-correlating each  
spectrum with a template spectrum formed from the average  
of 24 spectra from our run in 2000 December.   
The cross-correlation was made over the spectral region  
in the immediate vicinity of \ion{He}{1} $\lambda 6678$,  
although we also included the region surrounding  
\ion{He}{1} $\lambda 7065$ for the spectra from the  
2000 October and December runs that also recorded this line.  
These relative velocities were transformed to an absolute scale  
by adding the radial velocity of the template spectrum,   
$-43.1$ km~s$^{-1}$, found by parabolic fitting of the  
cores of \ion{He}{1} $\lambda\lambda 6678, 7065$.   
We also obtained cross-correlation velocities of the  
interstellar line at 6613 \AA ~to monitor the radial velocity  
stability of our measurements.  All these results are  
reported in Table~1 (given in full in the electronic  
edition) that lists the heliocentric Julian date  
of mid-exposure, orbital phase, radial velocity,  
observed minus calculated residual from the fit, and the  
interstellar line offset (plus columns describing the H$\alpha$ 
profile discussed in \S3).  The typical measurement error is  
$\pm 2.1$ km~s$^{-1}$ based upon the scatter between pairs  
of observations at closely spaced times.  
 
\placetable{tab1}      

Many recent papers on the orbital properties of LS~I~+61~303
adopt an arbitrary zero-phase at HJD 2,443,366.775 \citep{tay82}
and an orbital period of $26.4960\pm0.0028$~d, which was determined by 
\citet{gre02} from a careful study of the radio flux curve.  
Here we adopt the same convention for orbital phase that
we denote by $\phi ({\rm TG})$. 
 
We determined orbital elements using the non-linear, least-squares 
program of \citet{mor74}.  We also included  
the radial velocity data from the recent study by \citet{cas05}, 
but we found that our velocities are systematically lower than theirs 
by $-9.4$ km~s$^{-1}$ so we added this value to their measurements  
for ease of comparison.   Our first fit using all the available  
velocities is shown as the dotted line in Figure~1, and the resulting 
orbital elements are listed in column 4 of Table~2.  The r.m.s.\  
of this fit is 10.4 km~s$^{-1}$, which is about five times larger than  
the estimated measurement error.  This large scatter is caused by  
temporal changes in the shape of the \ion{He}{1} $\lambda 6678$  
profile.  For example, the six very low velocity measurements shown in  
Figure~1 were made during a remarkable episode of reduced H$\alpha$  
emission strength (\S3) when the \ion{He}{1} emission wings  
disappeared and the absorption core became broader.  
Thus, we selectively zero-weighted these six measurements and  
five other very discrepant points that also appeared to display  
profile shape changes.  The fit resulting from this edited set  
of velocities is shown as a solid line in Figure~1, and the  
corresponding orbital elements are given in column~5 of Table~2.  
 
\placetable{tab2}      
 
\begin{figure} 
\begin{center} 
{\includegraphics[angle=90,height=12cm]{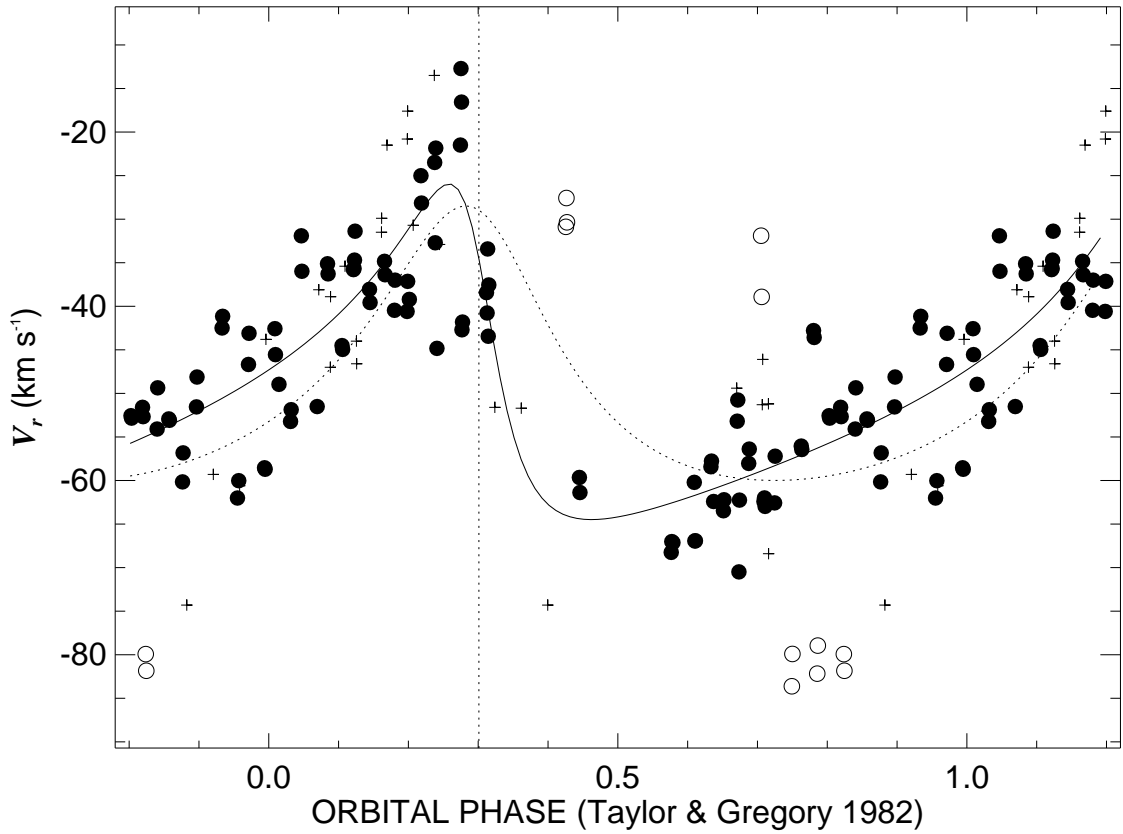}} 
\end{center} 
\caption{Calculated radial velocity curve ({\it solid line}) for  
LS~I~+61~303.  The measured radial velocities are shown as filled circles. 
Open circles mark the measurements that were assigned zero weight, and  
plus signs indicate the measurements from \citet{cas05}. 
The curved dotted line shows the orbital solution that results when  
the discrepant measurements are included in the fit.
The vertical dotted line indicates periastron phase.} 
\label{fig1} 
\end{figure}

\placefigure{fig1}     
 
Our results are generally in good agreement with earlier  
determinations of the elements by \citet{hut81} and \citet{cas05} 
(given in columns 2 and 3 of Table~2, respectively).  
The largest range in results is found for the orbital eccentricity  
$e$, and unfortunately the paucity of measurements in the  
orbital phase range 0.4 -- 0.5 limits the accuracy of the  
estimate in our work.  Our final result of $e=0.55 \pm 0.05$ 
is somewhat lower than the estimate of $e=0.8\pm0.1$ from  
fits of the IR light curve \citep{mar95}.   Models for the  
accretion flux variations indicate an eccentricity in the  
range $e=0.3$ \citep{lea01} to $e=0.6$ \citep{tay92}.    
Periastron occurs at phase $\phi ({\rm TG})=0.301\pm0.011$.  
This is in reasonable agreement with the expectation from the radio 
model of \citet{gre02} that places the phase of periastron in the  
range $\phi ({\rm TG})=0.33 - 0.40$.

 
\section{H$\alpha$ Variations}                      
 
The H$\alpha$ emission line is formed in the circumstellar disk of the  
Be star, and here we examine the temporal variations in the  
profile to investigate how the changes in the disk are related  
to changes in the accretion flux from the vicinity of the  
companion.  The H$\alpha$ profile appears as a double-peaked  
emission line in most of our spectra, and we made a number of  
measurements to characterize the shape and strength of the  
emission.  These measurements are summarized in Table~1.  
Column~6 of Table~1 lists the numerical integration  
of the line flux that includes the full extent of the  
line wings.  Since the emitting gas in the disk probably  
experiences Keplerian rotation \citep{por03}, the high velocity  
wings are formed closest to the Be star.  We measured the  
radial velocity of the wings based upon a bisector position  
determined using the method of \citet*{sha86}.  
This method samples the line wings using oppositely signed  
Gaussian functions and determines the mid-point position  
between the wings  
by cross-correlating these Gaussians with the profile.  
We used Gaussian functions with  
FWHM = 200 km~s$^{-1}$ at sample positions in the wings  
of $\pm390$ km~s$^{-1}$, and these radial velocities are  
given in column 7 of Table~1.   \citet{zam99} advocated  
making fits of the double-peaked H$\alpha$ profile using  
Gaussian functions to match the violet $V$ and red $R$ peaks,  
and we have followed their approach for consistency.   
These double-Gaussian fits were restricted to the inner  
part of the profile ($|\triangle \lambda | < 7$ \AA )  
since the wings are much more extended that those of 
Gaussian functions.   The remaining columns in Table~1  
list the parameters for these fits:  
radial velocity of the $V$ peak (col.~8),  
radial velocity of the $R$ peak (col.~9),  
ratio of the equivalent widths of the $V$ and $R$ components (col.~10), 
$V/R$ peak intensity ratio (col.~11),  
FWHM for the $V$ peak (col.~12), and  
FWHM for the $R$ peak (col.~13). 
 
We find evidence for temporal variations on timescales  
of days and longer.  The most dramatic change in the  
H$\alpha$ profile that we observed occurred in 1999 October. 
During the first three nights of this run the H$\alpha$  
profile appeared at normal strength (top plot in Fig.~2),  
but in the course of 24 hours or less the emission  
strength declined by about a factor of ten (lower plot in Fig.~2).  
The equivalent width over the next two days stayed at  
record low values relative to existing measurements  
\citep{par94,zam99,liu00,zam01,liu05}.  At the same time  
the emission in the wings of \ion{He}{1} $\lambda 6678$  
also disappeared revealing a broader photospheric line.  
\citet{liu05} obtained a spectrum six days after our  
last weak emission observation, and by that time the  
emission equivalent width was back to normal levels.  
We measured a projected rotational velocity 
for the weak emission spectra of 
$V\sin i = 104 \pm 5$ km~s$^{-1}$ from a comparison 
of the width of \ion{He}{1} $\lambda 6678$ with that in 
model profiles \citep{lan03} convolved with a rotational 
broadening function.  This is close to the value obtained 
by \citet{cas05}, but we caution that this is probably only 
a lower limit since the \ion{He}{1} profile (like H$\alpha$) 
is still probably affected by emission components even 
during this weak emission phase. 
 
\begin{figure} 
\begin{center} 
{\includegraphics[angle=90,height=12cm]{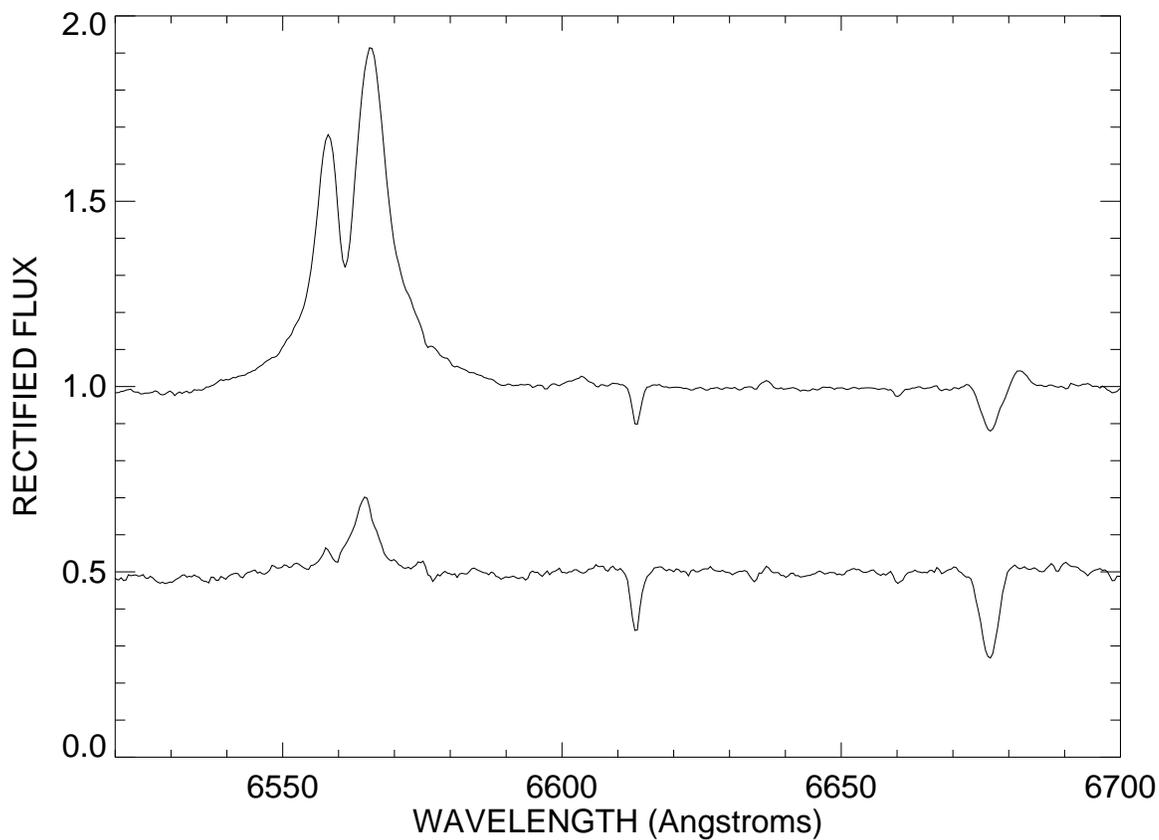}} 
\end{center} 
\caption{Average spectra of LS~I~+61~303 from 1999 October. 
The top plot shows the average from the first three nights when  
the H$\alpha$ emission strength was normal, while the bottom 
plot (offset by $-0.5$ in rectified flux for clarity) shows the  
average from the next three nights when the emission strength 
plummeted.  The other spectral features shown are the interstellar 
6613 \AA ~line and \ion{He}{1} $\lambda 6678$.} 
\label{fig2} 
\end{figure}

\placefigure{fig2}     
 
It is hard to imagine any process that would change the density  
properties of the Be star's disk so significantly in such  
a short time span.  We suggest instead that the neutral  
hydrogen in the disk was almost totally ionized at that  
time by exposure to energetic radiation from the vicinity  
of the companion.  We show in Figure~3 the contemporary  
variations of the radio flux as observed with the NRAO Green Bank  
Interferometer\footnote{ftp://ftp.gb.nrao.edu/pub/fghigo/gbidata/gdata/00README}  
\citep{ray97}, the daily averaged 5 -- 12 keV X-ray flux observed  
with {\it RXTE}/ASM\footnote{http://xte.mit.edu} \citep{lev96},  
and the H$\alpha$ equivalent width.  Although there is no 
evidence of any X-ray flare at that time, a secondary  
radio brightening did occur about the time of the H$\alpha$ decline.  
However, since the H$\alpha$ strength appeared relatively normal during  
other radio bursts, it is not clear how or if this secondary maximum  
is related to the H$\alpha$ weakening event.  If the radio emission  
forms in the relativistic jets, then it is possible that the secondary  
radio maximum originated at positions further away from the orbital plane 
where the inner Be star disk was more directly illuminated by jet emission.  
We note that the emission weakening event occurred near apastron,  
an orbital phase where \citet{liu05} also observed rapid variations 
in the H$\alpha$ profile.   
 
\begin{figure} 
\plotone{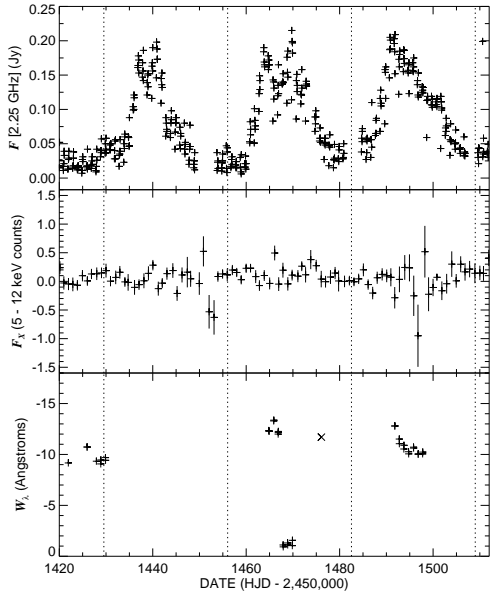}  
\caption{The time variations in radio flux ({\it top panel}),  
X-ray emission ({\it middle panel}), and H$\alpha$ emission  
equivalent width ({\it lower panel}) around the time of the  
H$\alpha$ emission decline.  The $X$ symbol in the lower plot 
represents an equivalent width measurement from \citet{liu05}. 
The dotted lines indicate times of periastron.} 
\label{fig3} 
\end{figure}

\placefigure{fig3}     
 
We next turn to H$\alpha$ variations related to the orbital period.  
\citet{zam99} demonstrated that the variations in the relative strengths  
of the $V$ and $R$ peaks are least partially related to orbital phase  
(see their Fig.~3), and our results appear to support their conclusion.  
We show in the lower panel of Figure~4 the ratio of the equivalent  
widths of the $V$ and $R$ fitted components (Table~1) plotted as  
a function of the orbital phase.  We see that the $V$ peak appears to  
strengthen relative to the $R$ peak around periastron, $\phi ({\rm TG})=0.30$.  
This strengthening of the short wavelength side of the profile also appears 
to shift the wing radial velocity to lower values.  The top panel of Figure~4 
shows how the wing velocity attains a minimum near periastron with a velocity  
curve that is markedly different from the orbital one (shown as a  
dotted line).  The middle panel shows the average radial velocity of  
the $V$ and $R$ peaks that show no evidence of orbital motion.   
 
\begin{figure} 
\plotone{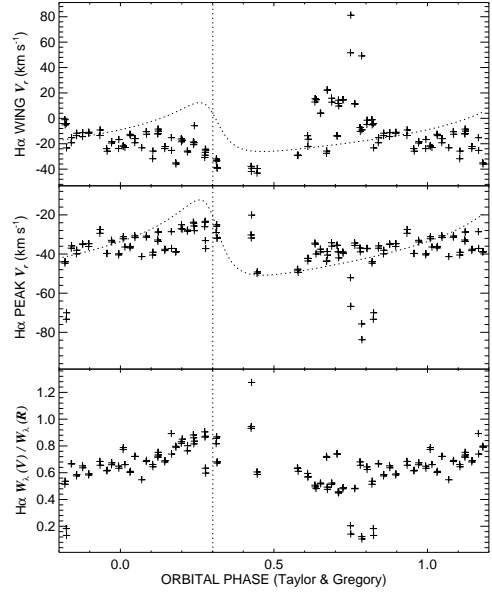}  
\caption{The orbital phase variations in the H$\alpha$ wing  
radial velocity ({\it top}), the mean H$\alpha$ peak radial  
velocity ({\it middle}), and the ratio of the equivalent  
widths of the $V$ and $R$ components ({\it bottom}).  
The curved dotted lines in the upper two panels show the orbital  
radial velocity curve translated to the mean velocity of  
the sample in each case.
The vertical dotted line indicates periastron phase.} 
\label{fig4} 
\end{figure}

\placefigure{fig4}     
 
We compare in Figure~5 the H$\alpha$ equivalent width variations with orbital phase  
to those observed in the radio and X-ray fluxes.  The lower panel shows our  
equivalent width measurements combined with 159 other published measurements 
\citep{par94,zam99,liu00,zam01,liu05} that we have binned into ten phase groups 
(after omitting the six unusually low measurements from the 1999 October decline).  
We see that when averaged over many cycles there appears to be a coherent  
cyclic variation in emission strength that attains a maximum near $\phi ({\rm TG})=0.55$.  
A similar kind of variation is seen in the orbital phase binned X-ray  
fluxes from {\it RXTE}/ASM \citep{par97,lea01}.  The top panel of Figure~5  
shows the binned radio curve from the Green Bank Interferometer.  
Note that the radio brightenings from cycle to cycle show significant  
variation in the shape of the light curve and phase of maximum \citep{gre02}.  
 
\begin{figure} 
\plotone{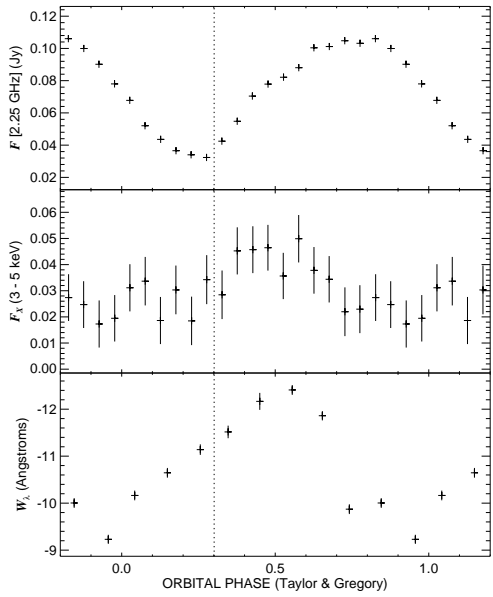}  
\caption{The variations in radio flux ({\it top panel}),  
X-ray emission ({\it middle panel}), and H$\alpha$ emission  
equivalent width ({\it lower panel}) binned and plotted as a  
function of orbital phase.  The vertical bars through each point 
express the standard deviation of the mean in the phase bin.
The vertical dotted line indicates periastron phase.} 
\label{fig5} 
\end{figure}

\placefigure{fig5}     
 
Finally we show in Figure~6 these same three fluxes over the period of  
active observation.  The top panel shows the Green Bank Interferometer 
flux density observations at 2.25 GHz that we averaged by binning into 
time intervals equal to three orbital periods (bins of 79.5~d), 
and the middle panel illustrates the {\it RXTE}/ASM X-ray fluxes binned 
in the same way.  The lower panel shows the individual H$\alpha$  
equivalent widths from our measurements ({\it plus signs}) and from  
other published measurements ({\it X signs}).  \citet{gre02} reported  
evidence of a long (1167~d) periodicity in the radio measurements,  
and \citet{zam00} showed that a comparable timescale is found in the  
H$\alpha$ properties.  There are two maxima visible that are separated by 
about 1200~d in the H$\alpha$ time series, but the coverage is too 
sparse to verify whether other cyclic maxima are present. 
   
\begin{figure} 
\plotone{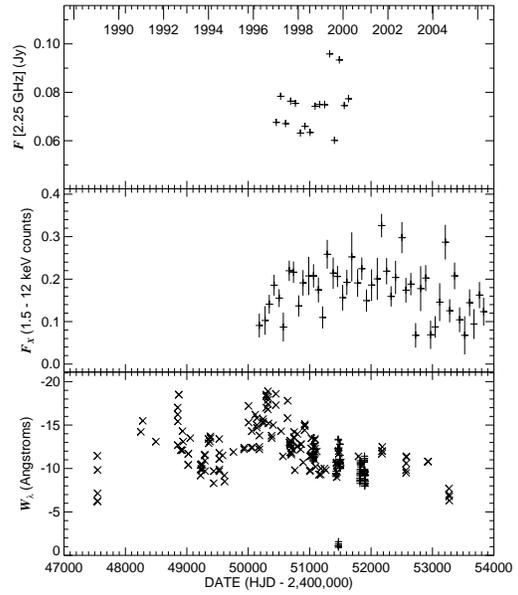}  
\caption{The time variations in radio flux ({\it top panel}),  
X-ray emission ({\it middle panel}), and H$\alpha$ emission  
equivalent width ({\it lower panel}) over the period of  
active observation.  The values in the upper two panels  
represent binned averages over time bins of three orbital  
periods.  The $X$ symbols in the lower plot 
represent other published equivalent width measurements.} 
\label{fig6} 
\end{figure}

\placefigure{fig6}     

 
\section{Discussion}                                
 
The circumstellar disks of the BeXRBs are comparable in many  
ways to those of other Be stars, but their outer radii are  
probably limited by the gravitational influence of their compact  
companions \citep{zam01}.  \citet*{rei97} show that the maximum  
H$\alpha$ emission equivalent widths (a good measure of the maximum  
disk size) are correlated with orbital period for the BeXRBs.  
\citet{oka01} found that these limiting radii are defined by the  
closest approach of the companion in the high eccentricity systems 
and by resonances between the orbital period and the disk gas  
rotational periods in the low eccentricity systems.   
LS~I~+61~303 has an orbital eccentricity that falls between these 
two cases, where the important resonance radii are comparable in  
size to the periastron separation. 
 
We can estimate the disk radius for LS~I~+61~303 from the 
H$\alpha$ equivalent width data shown in Figure~6.  
We recently presented numerical models of the circumstellar disks of Be 
stars \citep{gru06}, and we demonstrated that there are
monotonic relationships between the emission line equivalent
width and the ratio of the angular half-width at half maximum emission 
of the projected disk major axis to the radius of the star.
The relationship depends upon the temperature of the 
star, the inclination of the disk normal to the line of
sight, and the adopted outer boundary for the disk radius.
We assumed a primary star effective temperature of  
$T_{\rm eff} = 29850$~K appropriate to a B0~V classification 
\citep{har88}, although we note that a lower temperature may  
be possible \citep{how83}.  Figure~7 illustrates the relations  
for the inclination limits of $i=10^\circ$ and $i=60^\circ$  
suggested by \citet{cas05}, and we use the mean of these  
cases in the radius estimations.  Note that \citet{hut81} argue  
that the inclination may be larger, and if so, then our radius  
estimates will be somewhat low.  We adopted an outer boundary for 
the disk as the Roche radius of the primary star at apastron.  
This assumption is not critical to the radius estimate since only 
weak, optically thin emission originates in the outer parts of 
the disk model.  
 
\begin{figure} 
\begin{center} 
{\includegraphics[angle=90,height=12cm]{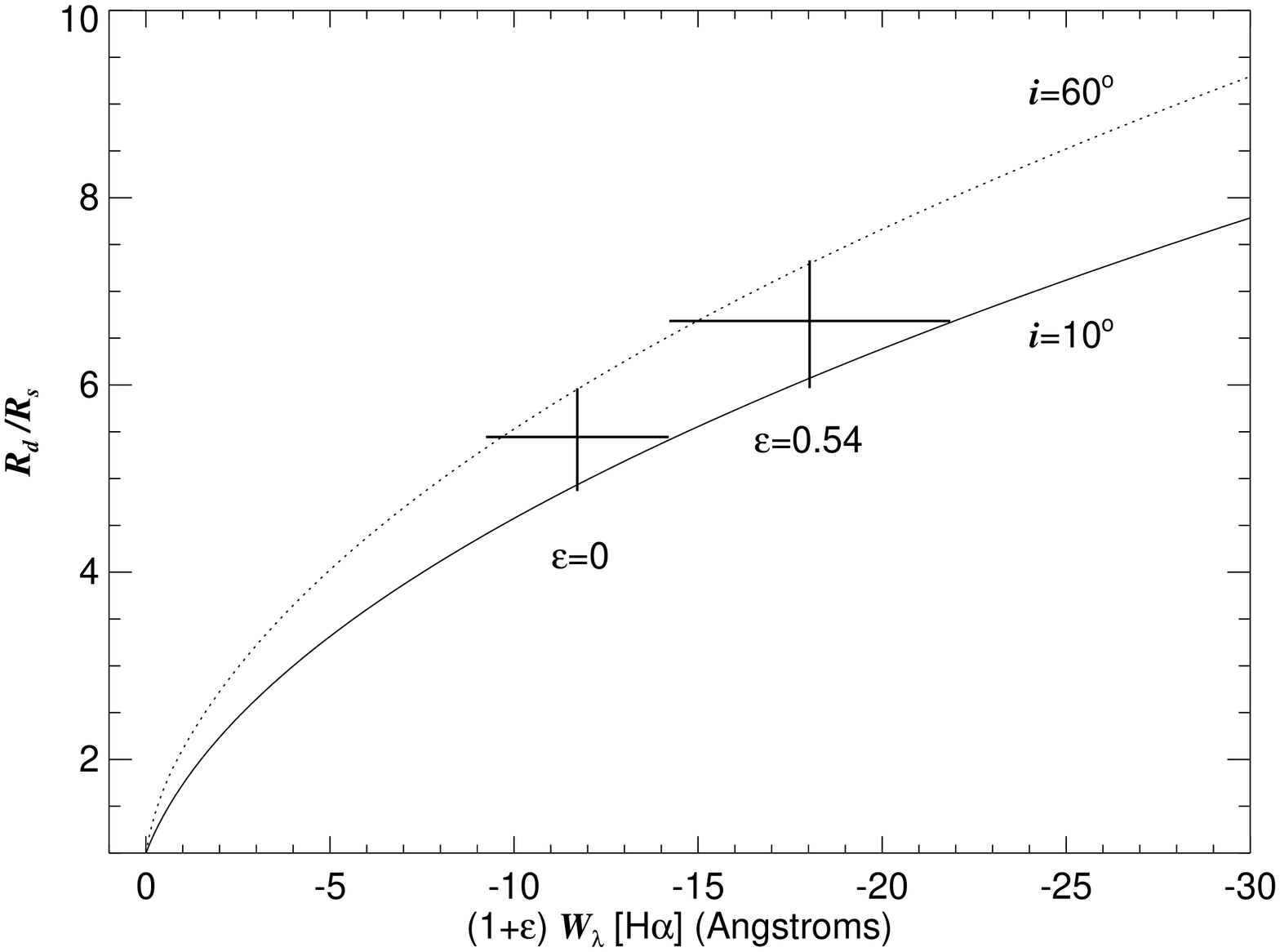}} 
\end{center} 
\caption{The predicted relationship between H$\alpha$ equivalent  
width (referenced to the stellar continuum level) and disk radius 
for two values of disk inclination.  The two crosses indicate  
the observed mean equivalent width with no adjustment for the  
disk continuum flux ($\epsilon=0$) and with the optical disk  
flux contribution suggested by \citet{cas05} ($\epsilon=0.54$).} 
\label{fig7} 
\end{figure}

\placefigure{fig7}    
 
The mean H$\alpha$ equivalent width from the data presented  
in Figure~6 (less the very low points from 1999 October) is 
$W_\lambda = -11.7 \pm 2.5$ \AA ~(standard deviation), and  
from the relations shown in Figure~7 the corresponding ratio of 
disk to stellar radius is $R_d/ R_s = 4.6\pm0.4$.   
The expected disk truncation radius due to resonances with  
the orbital motion is given by  
\begin{eqnarray} 
{R_n \over R_s} 
& = & \left({{GM_s}\over{4\pi^2}}\right)^{1/3} {1\over R_s} \left({P_{\rm orbit} \over n}\right)^{2/3} \\
& = & 37.4~ {{(M_s/M_\odot)^{1/3}}\over{R_s/R_\odot}}{1\over{n^{2/3}}} 
\end{eqnarray} 
where $n$ is the integer number of disk gas rotational periods  
per one orbital period.  \citet{oka01} predict that the $n=4$  
truncation radius will dominate in moderate eccentricity systems,  
and assuming a stellar mass of $M_s/M_\odot = 12.5 \pm 2.5$ \citep{cas05} 
and a radius of $R_s/R_\odot = 6.7\pm 0.9$ from the spectral  
classification \citep{har88}, then the predicted $n=4$ truncation  
radius is $R_4/R_s = 5.1 \pm 0.8$.  This agrees within errors  
with the estimate from the mean equivalent width given above, 
so that the observations are consistent with the idea that the  
disk radius is regulated by the $n=4$ resonance.  Note that  
\citet{cas05} suggest that the disk contributes about $35\%$ of  
the total continuum flux, so that the disk to stellar continuum  
flux ratio is $\epsilon= 0.54$ in the optical.  If so, then we need  
to rescale the equivalent width by a factor of $1+\epsilon$ in  
order to scale the H$\alpha$ emission flux to the stellar flux alone. 
This leads to an estimated disk radius of $R_d/R_s = 5.5\pm 0.5$  
(see Fig.~7), which is slightly larger than the $n=4$ truncation radius. 
 
The record of H$\alpha$ emission measurements indicates that the 
disk may occasionally grow to an even larger radius.  The two  
equivalent width maxima shown in Figure~6 reach $W_\lambda =  
-18.5\pm0.5$ \AA , and this value corresponds to a disk radius of  
$R_d/ R_s = 5.6 \pm 0.1$ (for $\epsilon = 0$).  The separation of 
the stars at periastron is given by  
\begin{equation} 
{R_p\over R_s} = {{(1-e) a} \over R_s} = 16.8~ {{(M_{\rm total}/M_\odot)^{1/3}} 
         \over {R_s/R_\odot}} 
\end{equation} 
and this is equal to $6.1\pm 1.2$ for $M_{\rm total}= (14.5\pm 4.0) M_\odot$ 
\citep{cas05}.  Thus, it appears that the times of the largest  
observed H$\alpha$ emission correspond to epochs when the disk  
almost reaches the limiting radius set by the close passage of the  
companion at periastron.   
 
\citet{oka02} have made numerical hydrodynamical simulations  
of the Be star's disk response to the gravitational influence  
of a compact companion in an elliptical orbit.  They show that the  
tidal pulls at periastron lead to the development of a large  
spiral wave in the disk that can extend far beyond the truncation  
radius and out to the vicinity of the companion (see their Fig.~11).  
Furthermore, \citet{hay05} show that a similar spiral pattern  
can be induced in the accretion disk surrounding the companion  
in the times shortly after periastron and that the presence of  
the pattern can promote mass accretion.  We speculate that  
processes like these are probably at work in LS~I~+61~303.  
The simulations show that the spiral wave starts at periastron  
as a density enhancement in the part of the disk facing the  
companion.  The longitude of periastron is $\omega = 57^\circ$,  
so the portion of the disk facing the companion at periastron will  
have a negative radial velocity for our line of sight, and the  
fact that we observe an increase in the strength of the approaching  
$V$ peak of H$\alpha$ at periastron (Fig.~4) is consistent with the  
formation of a spiral wave enhancement then.  As the spiral feature  
develops after periastron, disk gas would be carried outwards past  
the truncation radius, and the mass accretion processes would then  
begin in earnest.  The geometrical extension of the disk after  
periastron associated with the spiral feature would probably increase  
the projected area on the sky of high H$\alpha$ optical depth  
causing the emission feature to increase in strength as observed  
(Fig.~5).  
 
Our new determination of the time of periastron confirms  
that the X-ray and radio flux maxima occur some time after  
periastron \citep{mas04}.  Figure~5 shows that the X-ray flux  
attains a maximum near orbital phase $\phi {\rm (TG)}=0.5$ while the radio  
maximum occurs later near $\phi {\rm (TG)}=0.7$.  If the mass transfer process 
were instantaneous, we might expect that both fluxes would peak  
at periastron where the companion encounters the densest part of  
the disk.  However, \citet{tay92} and \citet{mar95} argue that  
if the process can be described by Bondi-Hoyle accretion then  
the mass accretion rate will also depend on the relative gas  
and accretor velocity and a second peak will occur later  
near apastron where the accretor's orbital velocity is slower.  
\citet{mas04} suggests that the first X-ray peak corresponds  
to the periastron accretion increase (with a phase lag  
caused by inverse Compton losses of relativistic electrons 
due to the large stellar flux field) and that the second 
radio peak occurs near the slow motion apastron phase  
(where Compton losses are reduced because the stellar flux  
is weaker).  If this radio flux peak arises from jets,  
then it is possible that ionizing radiation from the jets will lower 
the neutral H population and H$\alpha$ emission  
(leading to an H$\alpha$ minimum near apastron; see Fig.~5). 
We speculate that the ionizing flux may sometimes become quite strong  
(perhaps from knots that develop in the jets) and cause sudden decreases  
in H$\alpha$ emission strength such as we observed in 1999 October  
(Fig.~3).  
 
Our study indicates that the extent and shape of the Be star's 
disk has a significant influence on the mass transfer rate in this 
microquasar system.  We show in the next paper in this series 
that similar limiting disk radii and spiral wave extensions 
are also found in the BeXRB systems HDE~245770 and X~Per. 

 
\acknowledgments 
 
We thank Daryl Willmarth and the staff of KPNO for their assistance  
in making these observations possible.  The X-ray results were  
provided by the ASM/RXTE teams at MIT and at the RXTE SOF and GOF at NASA's GSFC. 
The Green Bank Interferometer is a facility of the National 
Science Foundation operated by the NRAO in support of  
NASA High Energy Astrophysics programs.   
This work was supported by the National Science Foundation under  
grants AST--0205297, AST--0506573, and AST--0606861. 
Institutional support has been provided from the GSU College 
of Arts and Sciences and from the Research Program Enhancement 
fund of the Board of Regents of the University System of Georgia, 
administered through the GSU Office of the Vice President 
for Research.   
 
 
\appendix

\section{Visual Companion}                          
 
LS~I~+61~303 has a nearby visual companion that we also  
recorded in the spectra made in 2000 December.  This companion 
star has a separation of $11\farcs2$ and a position angle of 
$167^\circ$ from LS~I~+61~303 according to coordinates  
listed in the Guide Star Catalog  
II\footnote{http://vizier.u-strasbg.fr/cgi-bin/VizieR?-source=I/271} 
(where the companion is designated 0404701738).  
The companion is approximately 3.5~mag fainter in the optical than  
LS~I~+61~303.  Our average spectrum shows a very broad H$\alpha$  
absorption line, no \ion{He}{1} $\lambda 6678$, a few weak  
metallic lines, and an interstellar line at 6613 \AA ~that is  
identical in appearance to its counterpart in the spectrum  
of LS~I~+61~303.  We compared the spectrum to several in the  
atlas by \citet{val04}, and we tentatively classify the star 
as type A2~V.  The radial velocity of the star from the H$\alpha$  
line is $-40\pm2$ km~s$^{-1}$.   
 
It is possible that this companion is an outlying member of the  
Cas~OB6 association as is LS~I~+61~303 \citep*{mir04}.  
The companion's radial velocity is comparable to that 
of other association members \citep{hil06} including  
the nearby multiple star HD~16429 \citep{mcs03}, and  
the similarity of the interstellar line strengths suggests  
a distance comparable to that of LS~I~+61~303. 
However, the companion also appears in the  
2MASS All-Sky Catalog of Point Sources \citep{cut03}   
(with a designation of 2MASS 02403208+6113340),  
and the near-IR magnitudes suggest that the star is  
closer than Cas~OB6.  Suppose the absolute $K$-band  
magnitude for a A2~V star is $M_K=1.2\pm0.6$ \citep{cox00} 
and that the extinction is $A(K)=0.36 E(B-V) = 0.3$  
\citep{how83,fit99}.  Then for the 2MASS magnitude of  
$K=11.55\pm0.02$, the estimated distance is $1.0\pm0.4$~kpc,  
which would place the star in the foreground of the  
Cas~OB6 stars at a distance of 1.9~kpc \citep{hil06}.   
Thus, unless the 2MASS magnitudes and/or the spectroscopic  
estimate of absolute magnitude are wrong, this companion 
is probably a chance optical alignment rather than  
a neighbor of LS~I~+61~303.   

 

 
\clearpage 
 
\begin{deluxetable}{lcccccccccccc} 
\rotate 
\tabletypesize{\scriptsize} 
\tablewidth{0pt} 
\tablecaption{Radial Velocity and H$\alpha$ Measurements\label{tab1}} 
\tablehead{ 
\colhead{Date} & 
\colhead{ } & 
\colhead{$V_r$} & 
\colhead{$(O-C)$}	& 
\colhead{$\triangle V_{\rm ISM}$}	& 
\colhead{$W_\lambda$}	& 
\colhead{$V_r (W)$} & 
\colhead{$V_r (V)$} & 
\colhead{$V_r (R)$} & 
\colhead{$W_\lambda (V)~/$} & 
\colhead{} & 
\colhead{FWHM($V$)} & 
\colhead{FWHM($R$)}  
\\ 
\colhead{(HJD$-$2,400,000)} & 
\colhead{$\phi ({\rm TG})$} & 
\colhead{(km s$^{-1}$)} & 
\colhead{(km s$^{-1}$)}	& 
\colhead{(km s$^{-1}$)} & 
\colhead{(\AA )} & 
\colhead{(km s$^{-1}$)}	& 
\colhead{(km s$^{-1}$)}	& 
\colhead{(km s$^{-1}$)} & 
\colhead{$W_\lambda (R)$} & 
\colhead{$V/R$} & 
\colhead{(\AA )} & 
\colhead{(\AA )}  
} 
\startdata 
 51053.860\dotfill &  0.122 & $ -34.7$ & \phn\phs$4.3$ & \phn   $-1.2$ &     $-$11.01 &  
\phn   $-8.5$ & $-202.5$ &     144.9 & 0.75 & 0.96 & 4.29 & 5.45 \\ 
 51053.881\dotfill &  0.123 & $ -31.4$ & \phn\phs$7.6$ & \phn   $-0.7$ &     $-$11.29 &  
      $-10.1$ & $-202.0$ &     144.5 & 0.73 & 0.95 & 4.21 & 5.43 \\ 
 51055.855\dotfill &  0.198 & $ -40.6$ & \phn   $-9.2$ & \phn\phs$1.6$ &     $-$11.65 &  
      $-16.2$ & $-197.8$ &     143.7 & 0.82 & 0.93 & 4.70 & 5.30 \\ 
 51055.877\dotfill &  0.199 & $ -37.1$ & \phn   $-5.9$ & \phn\phs$0.5$ &     $-$11.76 &  
      $-18.5$ & $-195.3$ &     145.3 & 0.83 & 0.89 & 4.77 & 5.14 \\ 
 51055.941\dotfill &  0.201 & $ -39.2$ & \phn   $-8.2$ & \phn\phs$2.1$ &     $-$11.75 &  
      $-17.2$ & $-197.9$ &     143.3 & 0.85 & 0.91 & 4.84 & 5.16 \\ 
 51056.902\dotfill &  0.237 & $ -23.5$ & \phn\phs$3.5$ & \phn\phs$0.3$ &     $-$11.29 &  
      $-18.6$ & $-196.3$ &     148.7 & 0.88 & 0.89 & 5.12 & 5.16 \\ 
 51056.923\dotfill &  0.238 & $ -32.7$ & \phn   $-5.8$ & \phn\phs$0.6$ &     $-$11.36 &  
      $-20.7$ & $-197.4$ &     149.0 & 0.86 & 0.88 & 5.05 & 5.19 \\ 
 51056.945\dotfill &  0.239 & $ -21.8$ & \phn\phs$5.0$ & \phn\phs$0.4$ &     $-$11.67 &  
      $-19.2$ & $-197.3$ &     146.0 & 0.82 & 0.91 & 4.85 & 5.40 \\ 
 51057.880\dotfill &  0.274 & $ -21.5$ & \phn\phs$5.5$ & \phn\phs$1.6$ &     $-$11.82 &  
      $-30.8$ & $-201.2$ &     149.1 & 0.91 & 0.90 & 5.37 & 5.33 \\ 
 51057.901\dotfill &  0.275 & $ -12.7$ &    \phs$14.4$ & \phn   $-0.4$ &     $-$11.21 &  
      $-29.1$ & $-196.0$ &     148.4 & 0.87 & 0.89 & 5.13 & 5.26 \\ 
 51057.922\dotfill &  0.276 & $ -16.6$ &    \phs$10.7$ & \phn   $-1.0$ &     $-$11.60 &  
      $-26.8$ & $-195.7$ &     149.2 & 0.87 & 0.90 & 5.20 & 5.36 \\ 
 51058.869\dotfill &  0.312 & $ -38.4$ & \phn\phs$1.0$ & \phn\phs$0.5$ &     $-$11.11 &  
      $-33.5$ & $-203.2$ &     145.2 & 0.82 & 0.90 & 5.09 & 5.60 \\ 
 51058.892\dotfill &  0.312 & $ -40.8$ & \phn   $-1.0$ & \phn\phs$1.6$ &     $-$10.80 &  
      $-32.1$ & $-201.8$ &     150.1 & 0.86 & 0.88 & 5.38 & 5.55 \\ 
 51058.913\dotfill &  0.313 & $ -33.4$ & \phn\phs$6.8$ & \phn   $-1.6$ &     $-$10.91 &  
      $-33.5$ & $-199.6$ &     149.4 & 0.87 & 0.91 & 5.29 & 5.55 \\ 
 51061.882\dotfill &  0.425 & $ -30.9$\tablenotemark{a} &    \phs$33.1$ & \phn   $-2.0$ &     $-$12.77 &  
      $-37.8$ & $-200.8$ &     140.3 & 0.93 & 0.92 & 6.53 & 6.48 \\ 
 51061.903\dotfill &  0.426 & $ -27.6$\tablenotemark{a} &    \phs$36.5$ & \phn   $-0.9$ &     $-$13.56 &  
      $-41.9$ & $-201.0$ &     137.3 & 0.95 & 0.93 & 6.63 & 6.54 \\ 
 51061.924\dotfill &  0.427 & $ -30.4$\tablenotemark{a} &    \phs$33.7$ & \phn   $-1.9$ &     $-$12.86 &  
      $-39.3$ & $-191.1$ &     150.8 & 1.27 & 0.99 & 7.60 & 5.89 \\ 
 51065.882\dotfill &  0.576 & $ -68.2$ & \phn   $-5.7$ & \phn\phs$1.0$ &     $-$11.35 &  
      $-29.1$ & $-225.8$ &     129.4 & 0.64 & 0.77 & 4.64 & 5.61 \\ 
 51065.903\dotfill &  0.577 & $ -67.0$ & \phn   $-4.4$ & \phn   $-0.3$ &     $-$11.30 &  
      $-29.1$ & $-226.2$ &     130.6 & 0.63 & 0.77 & 4.66 & 5.65 \\ 
 51065.946\dotfill &  0.579 & $ -67.1$ & \phn   $-4.6$ & \phn   $-0.6$ &     $-$11.65 &  
      $-29.2$ & $-225.2$ &     126.5 & 0.61 & 0.79 & 4.44 & 5.72 \\ 
 51066.756\dotfill &  0.609 & $ -60.2$ & \phn\phs$1.5$ & \phn   $-0.9$ &     $-$11.60 &  
      $-13.7$ & $-217.3$ &     132.9 & 0.59 & 0.76 & 4.67 & 5.96 \\ 
 51066.778\dotfill &  0.610 & $ -66.9$ & \phn   $-5.2$ & \phn\phs$1.0$ &     $-$12.30 &  
      $-22.1$ & $-215.8$ &     128.7 & 0.57 & 0.77 & 4.41 & 6.01 \\ 
 51066.800\dotfill &  0.611 & $ -66.9$ & \phn   $-5.2$ & \phn   $-1.3$ &     $-$12.24 &  
      $-16.5$ & $-214.4$ &     129.8 & 0.57 & 0.78 & 4.43 & 6.01 \\ 
 51421.937\dotfill &  0.014 & $ -49.0$ & \phn   $-2.5$ & \phn   $-0.2$ & \phn $-$9.16 &  
      $-23.0$ & $-209.5$ &     136.7 & 0.66 & 0.69 & 4.79 & 5.01 \\ 
 51425.943\dotfill &  0.165 & $ -34.8$ & \phn\phs$0.1$ & \phn   $-3.6$ &     $-$10.74 &  
      $-15.2$ & $-207.8$ &     150.7 & 0.89 & 0.92 & 5.60 & 5.75 \\ 
 51425.964\dotfill &  0.166 & $ -36.4$ & \phn   $-1.5$ & \phn   $-3.5$ &     $-$10.72 &  
      $-25.3$ & $-209.9$ &     135.6 & 0.74 & 0.82 & 4.95 & 5.50 \\ 
 51427.933\dotfill &  0.241 & $ -44.8$ &       $-18.1$ & \phn   $-1.3$ & \phn $-$9.34 &  
\phn   $-5.8$ & $-205.6$ &     149.5 & 0.84 & 0.85 & 6.20 & 6.32 \\ 
 51428.882\dotfill &  0.276 & $ -42.7$ &       $-15.4$ & \phn\phs$6.8$ & \phn $-$9.07 &  
      $-24.3$ & $-212.6$ &     146.4 & 0.63 & 0.69 & 5.17 & 5.60 \\ 
 51428.903\dotfill &  0.277 & $ -41.8$ &       $-14.4$ & \phn   $-0.8$ & \phn $-$9.46 &  
      $-25.1$ & $-214.2$ &     139.7 & 0.60 & 0.70 & 4.84 & 5.64 \\ 
 51429.880\dotfill &  0.314 & $ -43.4$ & \phn   $-2.9$ & \phn   $-1.2$ & \phn $-$9.69 &  
      $-38.7$ & $-202.4$ &     138.5 & 0.67 & 0.72 & 4.94 & 5.32 \\ 
 51429.901\dotfill &  0.315 & $ -37.5$ & \phn\phs$3.4$ & \phn   $-3.4$ & \phn $-$9.42 &  
      $-39.4$ & $-202.0$ &     138.3 & 0.68 & 0.75 & 4.90 & 5.41 \\ 
 51464.828\dotfill &  0.633 & $ -58.4$ & \phn\phs$2.6$ & \phn\phs$1.0$ &     $-$12.35 &  
   \phs$12.8$ & $-218.0$ &     149.1 & 0.51 & 0.71 & 4.35 & 6.09 \\ 
 51464.850\dotfill &  0.634 & $ -57.8$ & \phn\phs$3.3$ & \phn\phs$0.2$ &     $-$12.29 &  
   \phs$15.5$ & $-217.9$ &     148.0 & 0.49 & 0.70 & 4.33 & 6.13 \\ 
 51464.930\dotfill &  0.637 & $ -62.4$ & \phn   $-1.5$ & \phn   $-4.1$ &     $-$12.28 &  
   \phs$14.7$ & $-222.6$ &     142.7 & 0.48 & 0.69 & 4.31 & 6.18 \\ 
 51465.890\dotfill &  0.673 & $ -70.5$ &       $-10.6$ & \phn   $-3.6$ &     $-$13.29 &  
   \phs$22.4$ & $-220.5$ &     143.1 & 0.49 & 0.74 & 4.38 & 6.60 \\ 
 51465.911\dotfill &  0.674 & $ -62.2$ & \phn   $-2.4$ & \phn   $-3.2$ &     $-$13.37 &  
   \phs$22.1$ & $-222.1$ &     140.9 & 0.48 & 0.72 & 4.47 & 6.77 \\ 
 51466.840\dotfill &  0.709 & $ -62.4$ & \phn   $-3.6$ & \phn   $-3.7$ &     $-$12.00 &  
   \phs$11.7$ & $-221.2$ &     143.6 & 0.46 & 0.80 & 4.06 & 7.08 \\ 
 51466.861\dotfill &  0.710 & $ -62.0$ & \phn   $-3.2$ & \phn   $-3.8$ &     $-$12.16 &  
   \phs$14.3$ & $-220.4$ &     142.5 & 0.45 & 0.79 & 4.05 & 7.11 \\ 
 51466.884\dotfill &  0.711 & $ -63.0$ & \phn   $-4.2$ &       $-11.5$ &     $-$12.26 &  
\phn\phs$9.7$ & $-223.0$ &     139.1 & 0.45 & 0.79 & 4.05 & 7.04 \\ 
 51467.899\dotfill &  0.749 & $ -83.6$\tablenotemark{a} &       $-26.1$ &       $-14.8$ & \phn $-$1.11 &  
   \phs$51.6$ & $-200.5$ &      96.3 & 0.20 & 0.30 & 3.07 & 4.49 \\ 
 51467.920\dotfill &  0.750 & $ -79.9$\tablenotemark{a} &       $-22.4$ &       $-12.3$ & \phn $-$0.92 &  
   \phs$81.2$ & $-225.3$ &      91.9 & 0.14 & 0.20 & 3.28 & 4.70 \\ 
 51468.862\dotfill &  0.785 & $ -82.2$\tablenotemark{a} &       $-25.9$ & \phn   $-7.0$ & \phn $-$1.31 &  
   \phs$49.2$ & $-225.3$ &      74.0 & 0.12 & 0.35 & 2.11 & 6.02 \\ 
 51468.883\dotfill &  0.786 & $ -79.0$\tablenotemark{a} &       $-22.7$ & \phn   $-4.3$ & \phn $-$1.11 &  
\phn   $-9.1$ & $-246.5$ &      78.9 & 0.11 & 0.36 & 1.55 & 5.36 \\ 
 51469.872\dotfill &  0.823 & $ -79.9$\tablenotemark{a} &       $-25.0$ &       $-11.3$ & \phn $-$1.05 &  
\phn   $-4.2$ & $-230.1$ &      83.5 & 0.18 & 0.41 & 2.53 & 5.68 \\ 
 51469.893\dotfill &  0.824 & $ -81.8$\tablenotemark{a} &       $-26.9$ & \phn   $-8.0$ & \phn $-$1.57 &  
      $-23.3$ & $-226.9$ &      86.9 & 0.13 & 0.31 & 2.52 & 5.95 \\ 
 51491.798\dotfill &  0.651 & $ -63.5$ & \phn   $-2.9$ & \phn   $-5.3$ &     $-$12.80 &  
\phn\phs$4.0$ & $-220.5$ &     145.4 & 0.52 & 0.84 & 4.39 & 7.05 \\ 
 51491.820\dotfill &  0.652 & $ -62.2$ & \phn   $-1.7$ & \phn   $-4.0$ &     $-$12.79 &  
\phn\phs$4.0$ & $-222.4$ &     144.6 & 0.52 & 0.84 & 4.39 & 7.13 \\ 
 51492.757\dotfill &  0.687 & $ -58.0$ & \phn\phs$1.4$ & \phn\phs$0.1$ &     $-$11.51 &  
   \phs$15.8$ & $-221.3$ &     152.9 & 0.51 & 0.86 & 4.44 & 7.43 \\ 
 51492.779\dotfill &  0.688 & $ -56.4$ & \phn\phs$3.1$ & \phn\phs$0.5$ &     $-$11.06 &  
   \phs$12.7$ & $-224.6$ &     152.6 & 0.52 & 0.86 & 4.45 & 7.35 \\ 
 51493.740\dotfill &  0.724 & $ -62.6$ & \phn   $-4.3$ & \phn\phs$4.6$ &     $-$10.53 &  
   \phs$14.6$ & $-218.3$ &     140.2 & 0.48 & 0.92 & 3.82 & 7.28 \\ 
 51493.761\dotfill &  0.725 & $ -57.2$ & \phn\phs$1.1$ & \phn\phs$2.0$ &     $-$10.90 &  
   \phs$14.6$ & $-219.4$ &     140.1 & 0.49 & 0.93 & 3.87 & 7.38 \\ 
 51494.752\dotfill &  0.762 & $ -56.0$ & \phn\phs$1.0$ & \phn   $-2.5$ &     $-$10.27 &  
   \phs$11.6$ & $-211.5$ &     142.8 & 0.48 & 0.78 & 4.21 & 6.87 \\ 
 51494.773\dotfill &  0.763 & $ -56.4$ & \phn\phs$0.6$ & \phn\phs$0.8$ &     $-$10.06 &  
   \phs$11.2$ & $-213.0$ &     141.8 & 0.48 & 0.79 & 4.17 & 6.84 \\ 
 51495.809\dotfill &  0.802 & $ -52.6$ & \phn\phs$3.1$ & \phn   $-5.6$ &     $-$10.72 &  
\phn   $-4.8$ & $-212.9$ &     138.4 & 0.64 & 0.76 & 4.92 & 5.84 \\ 
 51495.830\dotfill &  0.803 & $ -52.8$ & \phn\phs$2.8$ & \phn   $-5.5$ &     $-$10.63 &  
\phn   $-1.5$ & $-211.4$ &     137.1 & 0.62 & 0.79 & 4.67 & 5.92 \\ 
 51496.800\dotfill &  0.840 & $ -54.1$ & \phn\phs$0.3$ & \phn   $-4.1$ &     $-$10.00 &  
      $-19.2$ & $-212.5$ &     138.5 & 0.67 & 0.75 & 5.12 & 5.77 \\ 
 51496.821\dotfill &  0.840 & $ -49.4$ & \phn\phs$5.0$ & \phn   $-0.1$ &     $-$10.08 &  
      $-15.2$ & $-209.6$ &     138.0 & 0.67 & 0.77 & 5.09 & 5.84 \\ 
 51497.765\dotfill &  0.876 & $ -60.2$ & \phn   $-7.2$ & \phn   $-1.3$ &     $-$10.06 &  
      $-14.4$ & $-215.1$ &     145.5 & 0.65 & 0.74 & 5.23 & 5.96 \\ 
 51497.786\dotfill &  0.877 & $ -56.8$ & \phn   $-3.9$ & \phn   $-3.5$ &     $-$10.23 &  
      $-11.2$ & $-216.8$ &     146.5 & 0.64 & 0.75 & 5.23 & 6.14 \\ 
 51817.802\dotfill &  0.955 & $ -62.0$ &       $-12.5$ & \phn   $-0.4$ & \phn $-$8.63 &  
      $-23.8$ & $-219.1$ &     139.6 & 0.61 & 0.72 & 4.21 & 4.92 \\ 
 51817.850\dotfill &  0.957 & $ -60.0$ &       $-10.6$ & \phn   $-0.1$ & \phn $-$8.53 &  
      $-25.9$ & $-218.1$ &     138.7 & 0.62 & 0.72 & 4.18 & 4.87 \\ 
 51818.830\dotfill &  0.994 & $ -58.5$ &       $-10.9$ & \phn\phs$1.2$ & \phn $-$8.85 &  
      $-23.8$ & $-221.4$ &     140.5 & 0.65 & 0.71 & 4.30 & 4.73 \\ 
 51818.851\dotfill &  0.994 & $ -58.7$ &       $-11.1$ & \phn\phs$0.8$ & \phn $-$8.29 &  
      $-16.7$ & $-221.7$ &     142.0 & 0.63 & 0.71 & 4.23 & 4.74 \\ 
 51819.817\dotfill &  0.031 & $ -53.2$ & \phn   $-7.7$ & \phn\phs$2.0$ & \phn $-$8.87 &  
      $-12.6$ & $-216.3$ &     142.6 & 0.60 & 0.79 & 4.35 & 5.68 \\ 
 51819.840\dotfill &  0.032 & $ -51.9$ & \phn   $-6.4$ & \phn   $-0.4$ & \phn $-$9.16 &  
      $-13.6$ & $-215.3$ &     143.0 & 0.61 & 0.72 & 4.53 & 5.33 \\ 
 51820.827\dotfill &  0.069 & $ -51.5$ & \phn   $-8.4$ & \phn\phs$2.9$ & \phn $-$9.30 &  
      $-23.2$ & $-215.9$ &     133.3 & 0.55 & 0.73 & 4.00 & 5.30 \\ 
 51821.770\dotfill &  0.105 & $ -44.5$ & \phn   $-4.0$ & \phn   $-0.6$ & \phn $-$9.51 &  
      $-31.8$ & $-212.4$ &     131.4 & 0.66 & 0.86 & 4.00 & 5.19 \\ 
 51821.792\dotfill &  0.105 & $ -45.0$ & \phn   $-4.5$ & \phn\phs$0.8$ & \phn $-$9.41 &  
      $-25.8$ & $-213.9$ &     135.7 & 0.64 & 0.86 & 3.98 & 5.29 \\ 
 51822.812\dotfill &  0.144 & $ -38.1$ & \phn   $-1.0$ & \phn\phs$1.9$ & \phn $-$9.91 &  
      $-23.8$ & $-212.5$ &     136.0 & 0.68 & 0.91 & 4.04 & 5.38 \\ 
 51822.834\dotfill &  0.145 & $ -39.6$ & \phn   $-2.5$ & \phn\phs$1.1$ & \phn $-$9.73 &  
      $-22.2$ & $-211.8$ &     136.7 & 0.69 & 0.93 & 4.00 & 5.38 \\ 
 51823.762\dotfill &  0.180 & $ -40.5$ & \phn   $-7.0$ & \phn   $-0.1$ &     $-$10.90 &  
      $-35.1$ & $-215.8$ &     137.7 & 0.80 & 0.93 & 4.60 & 5.35 \\ 
 51823.783\dotfill &  0.181 & $ -37.0$ & \phn   $-3.7$ & \phn\phs$2.0$ &     $-$10.75 &  
      $-36.0$ & $-213.7$ &     136.2 & 0.79 & 0.92 & 4.56 & 5.32 \\ 
 51824.765\dotfill &  0.218 & $ -25.0$ & \phn\phs$4.0$ & \phn\phs$5.7$ &     $-$10.59 &  
      $-25.8$ & $-204.2$ &     147.4 & 0.76 & 0.89 & 4.61 & 5.37 \\ 
 51824.786\dotfill &  0.218 & $ -28.1$ & \phn\phs$0.8$ & \phn\phs$3.4$ &     $-$10.43 &  
      $-26.5$ & $-202.6$ &     146.9 & 0.80 & 0.86 & 4.79 & 5.16 \\ 
 51830.779\dotfill &  0.445 & $ -59.6$ & \phn\phs$4.7$ & \phn\phs$0.4$ &     $-$10.44 &  
      $-43.1$ & $-221.7$ &     121.9 & 0.61 & 0.76 & 4.60 & 5.78 \\ 
 51830.801\dotfill &  0.445 & $ -61.4$ & \phn\phs$3.0$ & \phn\phs$1.4$ &     $-$10.18 &  
      $-39.5$ & $-220.5$ &     122.2 & 0.59 & 0.77 & 4.44 & 5.84 \\ 
 51889.762\dotfill &  0.671 & $ -53.2$ & \phn\phs$6.8$ & \phn\phs$1.2$ & \phn $-$9.43 &  
      $-27.5$ & $-219.8$ &     132.1 & 0.72 & 1.09 & 4.02 & 6.05 \\ 
 51889.783\dotfill &  0.671 & $ -50.8$ & \phn\phs$9.2$ & \phn   $-1.5$ & \phn $-$9.73 &  
      $-25.8$ & $-220.3$ &     133.3 & 0.71 & 1.09 & 3.96 & 6.04 \\ 
 51890.669\dotfill &  0.705 & $ -31.9$\tablenotemark{a} &    \phs$27.0$ & \phn   $-1.4$ & \phn $-$9.76 &  
      $-13.5$ & $-216.9$ &     145.8 & 0.74 & 1.13 & 4.03 & 6.14 \\ 
 51890.690\dotfill &  0.706 & $ -38.9$\tablenotemark{a} &    \phs$20.0$ & \phn\phs$0.2$ & \phn $-$9.54 &  
      $-14.2$ & $-215.2$ &     144.1 & 0.74 & 1.14 & 3.96 & 6.13 \\ 
 51892.658\dotfill &  0.780 & $ -42.8$ &    \phs$13.7$ & \phn   $-0.9$ &     $-$10.76 &  
\phn   $-7.3$ & $-216.1$ &     142.3 & 0.68 & 1.02 & 4.24 & 6.37 \\ 
 51892.682\dotfill &  0.781 & $ -43.6$ &    \phs$12.9$ & \phn   $-1.7$ &     $-$10.96 &  
      $-10.3$ & $-215.8$ &     138.0 & 0.66 & 1.00 & 4.20 & 6.39 \\ 
 51893.686\dotfill &  0.819 & $ -51.6$ & \phn\phs$3.5$ & \phn   $-2.4$ &     $-$11.15 &  
\phn   $-1.1$ & $-210.1$ &     120.8 & 0.51 & 0.89 & 3.90 & 6.80 \\ 
 51893.707\dotfill &  0.820 & $ -52.7$ & \phn\phs$2.4$ & \phn   $-0.4$ &     $-$11.42 &  
\phn   $-5.2$ & $-208.8$ &     121.1 & 0.54 & 0.91 & 3.91 & 6.66 \\ 
 51894.684\dotfill &  0.856 & $ -52.9$ & \phn\phs$0.8$ & \phn\phs$0.7$ & \phn $-$9.31 &  
      $-13.6$ & $-208.2$ &     128.4 & 0.58 & 0.84 & 4.02 & 5.86 \\ 
 51894.705\dotfill &  0.857 & $ -53.1$ & \phn\phs$0.6$ & \phn   $-0.8$ & \phn $-$9.35 &  
      $-11.8$ & $-206.7$ &     130.5 & 0.59 & 0.83 & 4.08 & 5.80 \\ 
 51895.733\dotfill &  0.896 & $ -51.5$ & \phn\phs$0.6$ & \phn\phs$0.7$ & \phn $-$9.88 &  
      $-11.8$ & $-203.9$ &     131.6 & 0.58 & 0.78 & 4.36 & 5.84 \\ 
 51895.754\dotfill &  0.897 & $ -48.1$ & \phn\phs$4.0$ & \phn   $-0.9$ & \phn $-$9.77 &  
      $-11.0$ & $-201.9$ &     132.5 & 0.59 & 0.79 & 4.25 & 5.68 \\ 
 51896.706\dotfill &  0.933 & $ -42.5$ & \phn\phs$8.1$ & \phn   $-0.4$ & \phn $-$8.73 &  
      $-13.5$ & $-203.7$ &     144.9 & 0.68 & 0.84 & 4.21 & 5.19 \\ 
 51896.727\dotfill &  0.934 & $ -41.1$ & \phn\phs$9.4$ & \phn   $-1.0$ & \phn $-$8.56 &  
\phn   $-9.2$ & $-201.7$ &     146.8 & 0.66 & 0.83 & 4.12 & 5.23 \\ 
 51897.707\dotfill &  0.971 & $ -46.7$ & \phn\phs$2.1$ & \phn\phs$0.2$ & \phn $-$8.02 &  
      $-19.4$ & $-206.4$ &     140.4 & 0.66 & 0.89 & 3.77 & 5.08 \\ 
 51897.728\dotfill &  0.971 & $ -43.1$ & \phn\phs$5.7$ & \phn   $-1.1$ & \phn $-$7.97 &  
      $-18.3$ & $-206.7$ &     139.0 & 0.67 & 0.90 & 3.83 & 5.11 \\ 
 51898.711\dotfill &  0.008 & $ -42.6$ & \phn\phs$4.3$ & \phn   $-0.7$ & \phn $-$8.24 &  
      $-21.9$ & $-207.4$ &     143.1 & 0.77 & 0.92 & 4.23 & 5.04 \\ 
 51898.732\dotfill &  0.009 & $ -45.5$ & \phn\phs$1.2$ & \phn   $-0.6$ & \phn $-$8.28 &  
      $-21.8$ & $-206.7$ &     144.4 & 0.79 & 0.91 & 4.34 & 5.00 \\ 
 51899.713\dotfill &  0.046 & $ -31.9$ &    \phs$12.7$ & \phn\phs$0.2$ & \phn $-$8.52 &  
      $-19.2$ & $-207.1$ &     145.7 & 0.72 & 0.89 & 4.24 & 5.26 \\ 
 51899.734\dotfill &  0.047 & $ -36.0$ & \phn\phs$8.6$ & \phn   $-0.1$ & \phn $-$8.54 &  
      $-16.0$ & $-208.2$ &     145.0 & 0.72 & 0.90 & 4.26 & 5.29 \\ 
 51900.708\dotfill &  0.084 & $ -35.1$ & \phn\phs$7.0$ & \phn\phs$2.0$ & \phn $-$9.23 &  
      $-10.5$ & $-200.3$ &     137.5 & 0.68 & 0.90 & 4.21 & 5.52 \\ 
 51900.729\dotfill &  0.085 & $ -36.3$ & \phn\phs$5.7$ & \phn\phs$1.0$ & \phn $-$9.56 &  
      $-12.5$ & $-200.7$ &     139.0 & 0.69 & 0.90 & 4.24 & 5.52 \\ 
 51901.693\dotfill &  0.121 & $ -35.8$ & \phn\phs$3.4$ & \phn\phs$0.7$ & \phn $-$9.18 &  
      $-12.3$ & $-200.1$ &     134.4 & 0.72 & 0.92 & 4.23 & 5.36 \\ 
 51901.714\dotfill &  0.122 & $ -35.6$ & \phn\phs$3.4$ & \phn\phs$1.3$ & \phn $-$9.63 &  
      $-11.9$ & $-201.5$ &     134.3 & 0.72 & 0.88 & 4.34 & 5.34 \\ 
\enddata 
\tablenotetext{a}{Assigned zero weight.} 
\end{deluxetable} 
 
\clearpage 
 
 
\begin{deluxetable}{lcccc} 
\rotate 
\tablewidth{0pc} 
\tablecaption{Orbital Elements\label{tab2}} 
\tablehead{ 
\colhead{Element}	&  
\colhead{\citet{hut81}}	& 
\colhead{\citet{cas05}}	& 
\colhead{All Measures}	& 
\colhead{Edited Set}	} 
\startdata 
$P$\tablenotemark{a} (d) \dotfill & 26.51          & 26.4960        & 26.4960        & 26.4960           \\ 
$T$ (HJD--2,400,000)     \dotfill & $43559 \pm 1$  & $43372.9\pm0.5$& $51057\pm1$    & $51058.6 \pm 0.3$ \\ 
Periastron ($\phi({\rm TG})$)\dotfill&$0.25\pm0.04$& $0.23\pm0.02$  & $0.24\pm0.04$  & $0.301\pm0.011$   \\   
$e$                      \dotfill & $0.60\pm 0.13$ & $0.72 \pm 0.15$& $0.34 \pm 0.08$& $0.55 \pm 0.05$   \\ 
$\omega$ (deg)           \dotfill & $46 \pm 23$    & $21 \pm 13$    & $15 \pm 17$    & $57 \pm 9$        \\ 
$K$ (km s$^{-1}$)        \dotfill & $24 \pm 6$     & $23 \pm 6$     & $15.8 \pm 1.4$ & $19.3 \pm 1.5$    \\ 
$\gamma$ (km s$^{-1}$)   \dotfill & $-57 \pm 3$    & $-40.2 \pm 1.9$& $-49 \pm 1$    & $-51.0 \pm 0.8$   \\ 
$f(m)$ ($M_\odot$)       \dotfill & $0.019\pm0.015$& $0.011\pm0.012$& $0.009\pm0.003$& $0.011\pm0.003$   \\ 
$a_1 \sin i$ ($R_\odot$) \dotfill & $10 \pm 3$     & $8 \pm 3$      & $7.8\pm0.7$    & $8.4 \pm 0.8$     \\ 
r.m.s. (km s$^{-1}$)     \dotfill & 15.6           & 8.5            & 10.4           & 7.5               \\ 
\enddata 
\tablenotetext{a}{Fixed.} 
\end{deluxetable} 
 
 
 
\clearpage 
 
\input{epsf} 
 
\clearpage 
 
\clearpage 
 
\clearpage 
 
\clearpage 
 
\begin{thebibliography}{} 
\bibitem[Albert et al.(2006)]{alb06} 
         Albert, J., et al. 2006, Science, 312, 1771
\bibitem[Casares et al.(2005)]{cas05} 
         Casares, J., Ribas, I., Paredes, J. M., Mart\'{i}, J., \&  
         Allende Prieto, C. 2005, \mnras, 360, 1105	 
\bibitem[Coe(2000)]{coe00} 
         Coe, M. J. 2000, in The Be Phenomenon in Early-Type Stars,  
         IAU Coll. 175 (ASP Conf. Vol. 214), ed. M. A. Smith,  
         H. F. Henrichs, \& J. Fabregat (San Francisco: ASP), 656 
\bibitem[Cox(2000)]{cox00} 
         Cox, A. N. (ed.) 2000, Allen's Astrophysical Quantities (4th Ed.) 
         (New York: AIP/Springer-Verlag) 
\bibitem[Cutri et al.(2003)]{cut03} 
         Cutri, R. M., et al. 2003, The 2MASS All-Sky Catalog of Point Sources 
         (Pasadena: Univ. Massachusetts/IPAC) 
\bibitem[Fitzpatrick(1999)]{fit99} 
         Fitzpatrick, E. L. 1999, \pasp, 111, 63 
\bibitem[Gies et al.(2003)]{gie03} 
         Gies, D. R., et al. 2003, \apj, 583, 424 
\bibitem[Gregory(2002)]{gre02} 
         Gregory, P. C. 2002, \apj, 575, 427 
\bibitem[Grundstrom \& Gies(2006)]{gru06} 
         Grundstrom, E. D., \& Gies, D. R. 2006, \apjl, in press 
\bibitem[Grundstrom et al.(2006a)]{gr06a} 
         Grundstrom, E. D., et al. 2006a, \apj, submitted  
\bibitem[Grundstrom et al.(2006b)]{gr06b} 
         Grundstrom, E. D., et al. 2006b, \apj, submitted   
\bibitem[Harmanec(1988)]{har88} 
         Harmanec, P. 1988, Bull. Astr. Inst. Cz., 39, 329 
\bibitem[Hayasaki \& Okazaki(2005)]{hay05} 
         Hayasaki, K., \& Okazaki, A. T. 2005, \mnras, 360, L15 
\bibitem[Hillwig et al.(2006)]{hil06} 
         Hillwig, T. C., Gies, D. R., Bagnuolo, W. G., Jr., Huang, W.,  
         McSwain, M. V., \& Wingert, D. W. 2006, \apj, 639, 1069 
\bibitem[Howarth(1983)]{how83} 
         Howarth, I. D. 1983, \mnras, 203, 801 
\bibitem[Hutchings \& Crampton(1981)]{hut81} 
         Hutchings, J. B., \& Crampton, D. 1981, \pasp, 93, 486 
\bibitem[Lanz \& Hubeny(2003)]{lan03}
         Lanz, T., \& Hubeny, I. 2003, \apjs, 146, 417	
\bibitem[Leahy(2001)]{lea01} 
         Leahy, D. A. 2001, \aap, 380, 516  
\bibitem[Levine et al.(1996)]{lev96} 
         Levine, A. M., Bradt, H., Cui, W., Jernigan, J. G., Morgan, E. H., 
         Remillard, R., Shirey, R. E., \& Smith, D. A. 
         1996, \apj, 469, L33 
\bibitem[Liu et al.(2000)]{liu00} 
         Liu, Q. Z., Hang, H. R., Wu, G. J., Chang, J., \& Zhu, Z. X. 
         2000, \aap, 359, 646 
\bibitem[Liu \& Yan(2005)]{liu05} 
         Liu, Q. Z., \& Yan, J. Z. 2005, New Astr., 11, 130 
\bibitem[Massi(2004)]{mas04} 
         Massi, M. 2004, \aap, 422, 267 
\bibitem[Massi et al.(2004)]{mea04} 
         Massi, M., Rib\'{o}, M., Paredes, J. M., Garrington, S. T.,
         Peracaula, M., \& Mart\'{i}, J. 2004, \aap, 414, L1
\bibitem[Mart\'{i} \& Paredes(1995)]{mar95} 
         Mart\'{i}, J., \& Paredes, J. M. 1995, \aap, 298, 151 
\bibitem[McSwain(2003)]{mcs03} 
         McSwain, M. V. 2003, \apj, 595, 1124 
\bibitem[McSwain et al.(2004)]{mcs04} 
         McSwain, M. V., Gies, D. R., Huang, W., Wiita, P. J.,  
         Wingert, D. W., \& Kaper, L. 2004, \apj, 600, 927 
\bibitem[Mirabel et al.(2004)Mirabel, Rodrigues, \& Liu]{mir04} 
         Mirabel, I. F., Rodrigues, I., \& Liu, Q. Z.  
         2004, \aap, 422, L29 
\bibitem[Morbey \& Brosterhus(1974)]{mor74} 
         Morbey, C., \& Brosterhus, E. B. 1974,  
        \pasp, 86, 455 
\bibitem[Okazaki et al.(2002)]{oka02} 
         Okazaki, A. T., Bate, M. R., Ogilvie, G. I., \& Pringle, J. E. 
         2002, \mnras, 337, 967 
\bibitem[Okazaki \& Negueruela(2001)]{oka01} 
         Okazaki, A. T., \& Negueruela, I. 2001, \aap, 377, 161 
\bibitem[Paredes et al.(1997)]{par97} 
         Paredes, J. M., Mart\'{i}, J., Peracaula, M., \& Ribo, M. 
         1997, \aap, 320, L25 
\bibitem[Paredes et al.(1994)]{par94} 
         Paredes, J. M., et al. 1994, \aap, 288, 519 
\bibitem[Porter \& Rivinius(2003)]{por03} 
         Porter, J. M., \& Rivinius, Th. 2003, \pasp, 115, 1153 
\bibitem[Punsly(1999)]{pun99} 
         Punsly, B. 1999, \apj, 519, 336 
\bibitem[Ray et al.(1997)]{ray97} 
         Ray, P. S., et al. 1997, \apj, 491, 381 
\bibitem[Reig et al.(1997)Reig, Fabregat, \& Coe]{rei97} 
         Reig, P., Fabregat, J., \& Coe, M. J. 1997, \aap, 322, 193  
\bibitem[Shafter et al.(1986)Shafter, Szkody, \& Thorstensen]{sha86} 
         Shafter, A. W., Szkody, P., \& Thorstensen, J. R. 1986, \apj, 308, 765 
\bibitem[Taylor \& Gregory(1982)]{tay82}
         Taylor, A. R., \& Gregory, P. C. 1982, \apj, 255, 210
\bibitem[Taylor et al.(1992)]{tay92} 
         Taylor, A. R., Kenny, H. T., Spencer, R. E., \& Tzioumis, A. 
         1992, \apj, 395, 268 
\bibitem[Wen et al.(2006)]{wen06} 
         Wen, L., Levine, A. M., Corbet, R. H. D., \& Bradt, H. V. 
         2006, \apjs, 163, 372 
\bibitem[Valdes et al.(2004)]{val04} 
         Valdes, F., Gupta, R., Rose, J. A., Singh, H. P., \& Bell, D. J. 
         2004, \apjs, 152, 251 
\bibitem[Zamanov \& Mart\'{i}(2000)]{zam00} 
         Zamanov, R. K., \& Mart\'{i}, J. 2000, \aap, 358, L55 
\bibitem[Zamanov et al.(1999)]{zam99} 
         Zamanov, R. K., Mart\'{i}, J., Paredes, J. M., Fabregat, J., 
         Rib\'{o}, M., \& Tarasov, A. E. 1999, \aap, 351, 543 
\bibitem[Zamanov et al.(2001)]{zam01} 
         Zamanov, R. K., Reig, P., Mart\'{i}, J., Coe, M. J.,  
         Fabregat, J., Tomov, N. A., \& Valchev, T. 
         2001, \aap, 367, 884	 
\end{thebibliography}
\end{document}